# Transforming building industry and health outcomes through social data-supported design

Melissa Marsh, Assoc. AIA
PLASTARC
New York, NY, USA
melissa@plastarc.com

Ingrid Erickson, Ph.D
Rutgers University
New Brunswick, NJ, USA
ingrid.erickson@rutgeters.edu

Jonah Bleckner
PLASTARC
New York, NY, USA
jonah@plastarc.com

**ABSTRACT**
A glaring reality of American industrialized society is that people spend a tremendous amount of their waking life in their workplace and other interior environments. Despite the amount of time that we spend in them, many of our constructed environments that we inhabit are not designed for the people and communities that rely on them. From a health and wellness perspective, there is a growing body of research on the ways that our interior environments and lack of exposure to natural elements systematically impacts our health and strains our healthcare system. In short, the spaces in which we live and work are a major public health issue and should be considered in this way. In this paper, we lay out a vision for using the underleveraged social data–available through social media–to inform the architecture, developer and real estate industries. The goal is ultimately a public health initiative: to create spaces that are healthier, more responsive, equitable and human-centric through social data-supported design.

## 1. INTRODUCTION

A number of statistical findings have highlighted the fact that workers in industrialized societies spend a tremendous amount of their waking life at their workplace. According to the American Time Use Survey, full-time workers spend an average of 8.7 hours per workday at the workplace, which cumulatively is an average of 1,790 hours per year. To put this figure in perspective, this translates to working Americans spending more time in their indoor office environments than eating, sleeping or participating in any leisure activities. The health impact of our architectural spaces appears even more profound if we consider that Americans, according to a National Human Activity Pattern Survey, spend roughly 93% of their time in buildings and automobiles. In this light, it is not an exaggeration to claim that our architectural spaces shape who we are simply as a function of the amount of time that we spend within and around them.

In addition to the numbers that have driven a realization about interior spatial occupancy, there are also a number of findings and forecasts that have highlighted the extreme public health impact of the cumulative time we spend in our built environments. For example, according to an estimate published in the *Journal of the American Medical Association*, the U.S spends around $86 billion dollars per year diagnosing and treating back and neck pain that is often caused from sitting at desks for prolonged periods of the workday. And it is no secret that the workplace supports the activity of sitting very well. A recent study published by the *International Journal of Behavioral Nutrition and Physical Activity* reported that people spend an average of 64 hours per week sitting. From a public health perspective, spaces like the modern workplace are the epitome of inactive design that normalize unhealthy habits rather than challenge them physically and socially.

The realization that our built environment is crucial to our health has provoked shifts in the architectural and real estate industries such that there is an emerging real estate value proposition for occupant-centric design. Proof of this industry shift is the upcoming institutionalization of Delos Living's WELL Building Standard, a performance-based certification system for evaluating the impact of the built environment on human health and well-being. The emphasis being placed on user experience-driven design has been further reinforced by the era of big data. At this moment of increasing data availability, transparency and optimization in other fields, architects are being asked to bring evidence that our solutions can and will perform for a variety of occupant objectives. However, a number of tactical questions must be addressed in order to begin to bridge the gap between the ideal and the reality of incorporating complex data into the spatial design and building development process.

For this paper, we propose an answer to one of the crucial questions surrounding data-driven design; at a time when we are inundated with information, where do we look for the data that will drive architectural designs that better meet our human and social needs? We advocate for using data available through social media platforms to understand how occupants feel about the spaces that they use. From this sentiment data, architects can then incorporate occupant feedback into their design approach in order to create spaces that are designed by us and for us. The power of social media, we find, is that it has the capacity to impact both space and behavior.

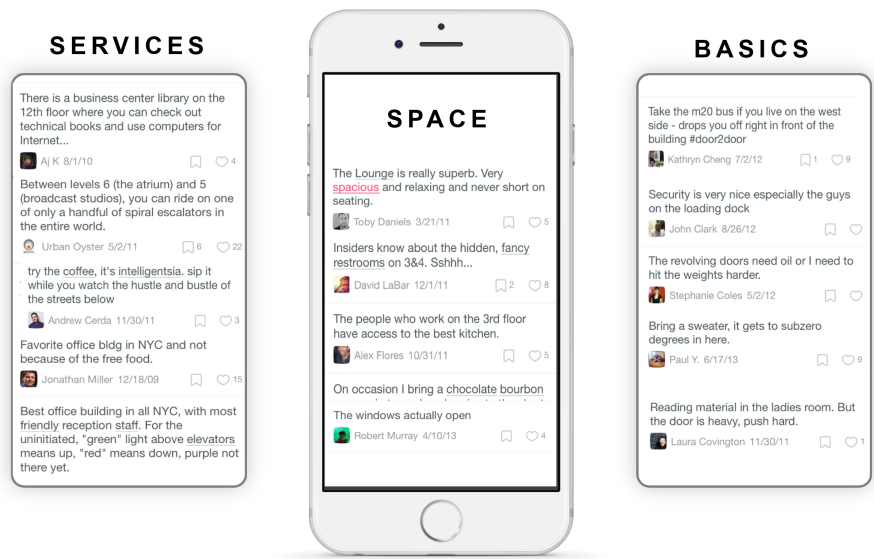

Snapshot of workplace feedback on rating and review app, Foursquare

## 2. WHERE IS THE DATA THAT MATTERS?

There are four general categories of data types that are essential for understanding the relationship between space and people. In this section, we introduce these four categories briefly in order of increasing complexity, and then describe where within the scheme social media data fits.

The first category is *location data*. The capacity to understand where people locate themselves is important information that can be leveraged by designers. The degree that a space is occupied helps designers determine what spatial features are either desirable or problematic. Some of the tools that fit neatly into this category include: zip code analyses, observational studies, commute analyses and badge swipe data analyses.

The second category that we have identified is *activity data*. By this we mean information that describes what people are doing. Are they sitting or standing? Are they working at a computer? Are they doing an active or inactive activity? These are a number of the questions that we ask in order to collect activity data.

Third, there is *performance data* that aims to record how an environment contributes to a person's activity and disposition. For example, in the world of workplace strategy, this includes tracking productivity-related metrics such as minutes lost per day, perceived level of productivity and other business measures. More generally though, performance data also includes health indicators and biometric data.

Finally, there is *sentiment data*, which intends to encapsulate how people feel about the environment; this is crucial because it is the holy grail of user experience design. Historically, sentiment data has been collected through social research tools like interviews, surveys and focus groups.

Beyond these four categories, a fifth category of data types emerges that is the product of combining and overlapping location, activity, performance and sentiment data in order to make new discoveries about the relationship between a space and its occupants. For example, the reason that location data is valuable from the perspective of a designer is because it is embedded with occupant sentiment information; the assumption is that if someone likes a feature of a space, they will occupy it more frequently. Furthermore, sentiment data, in the form of a social media post may also be embedded with location information. The possible intersections and hybrids of these four data types are endless.

Beyond the traditional social research tools, we now have access to another tool that should be leveraged to collect and analyze sentiment data related to the built environment: social media platforms. One example of the untapped potential of social media as a repository of human factors data is illustrated in a workplace research study conducted by PLASTARC. As part of the research, PLASTARC queried the Foursquare API to investigate how people are using social media to talk about their workplace environments. Their analysis found that an impressive array of occupants from the intern to the workplace community manager was using Foursquare to comment on a breadth of topics. In this sense, Foursquare was a stage where people were engaging in conversations about their relationship to the workplace: including occupant feedback about the entire workplace experience from building security and elevator speed, to coffee quality, design, aesthetic, technology, and even the bathroom.

Historically, commentary and feedback about the built environment has been a professionalized discipline reserved for architects and architectural critics–as well as the patrons with the capital to pay for the architect– now with platforms like Twitter, Instagram and Foursquare any user can comment on the way that a space feels, looks, performs, and so on. In this sense, social media has the capacity to contribute to the democratization of the design process and outcomes by expanding the boundaries of participation. Since we all spend a lot of time in architectural spaces, we should also all have a stake in the design process both through direct participation and data collection. Social media has the capacity to enable the production of spaces that are built by us and for us.

## 3. CHALLENGES TO IMPLEMENTATION

Tracking spaces that are desired, sought after, bragged about, and checked into presents an exciting opportunity to further understand the capability and power of our design solutions. The urban design and planning professional communities provide a positive comparative example in leveraging new public data sets to support innovative solutions for communities. Consumer research has also leveraged, albeit sometimes controversially, consumer data to predict everything from shopping behavior to product preference. As other disciplines increasingly leverage data and user experience, now is the time for architects and building industry professionals to be educated and aware, seek out, and incorporate human factors data as they forge a future built environment that serves the needs of an intelligible occupant.

The reality is that big data in the eye of an architect can be difficult to decipher without the familiarity, tools, and language that other disciplines have comprised. The current mismatch of big data rhetoric and the architect's desire to 'see and feel' is an opportunity to re-introduce one to the other. The success and continued improvement of the architectural discipline requires that architects be able to work with different sets of data, bring these influencers into the existing rhetoric, and embed a curiosity to explore more novel data sources for designing and building new initiatives. Data should not control, but be included as one of the many informants to the architect's work. The social data that exists through social media is one such source of information that can inform a more holistic and human-centric design approach. Furthermore, the benefits of mining intelligence from social data are not isolated to the architecture professional community. Many communities stand to benefit when social media helps buildings better understand occupants and occupants better understand buildings.

## 4. CONCLUSION

We as individuals and communities are shaped by the architectural spaces we build. Not only does the built environment communicate and reproduce cultural values and ideals, there is a growing body of research that shows that these spaces also impact human health. That being said, architecture and design are not sufficiently considered a matter of public health concern. In this paper, we argue that our built environment is precisely that: a critical public health challenge. However, the design of the built environment is also a tool that can be strategically deployed to combat a lot of the systematic public health challenges that communities face. In order to leverage this potential of architecture and design, we proposed in this paper that designers utilize the wealth of sentiment and other types of data that are made available through social media platforms in order to implement a more participatory and human-centric approach to spatial designs. Furthermore, it is important to recognize that the use of social data-supported design is not exclusively a tool to regulate and avoid bad outcomes, but also a significant investment in optimized and good high performance physical environments. Through social media, people are already talking about the environments they inhabit. The crucial question is whether you are paying enough attention?

## 5. ACKNOWLEDGEMENTS

Thank you to the entire PLASTARC team for the on-going research and conversations that form the foundation of this paper; Thanks to Claire Rowell for her dedication to enabling every member of the PLASTARC team; to Kristin Mueller for always lifting team spirits with the perfect recipe of smiles, songs and sweets; to Cassie Hackel and Roger Marsh, who are more than a voice that you can always count on.

Finally, thank you to all the researchers and practitioners that continue to challenge and questions the institutional boundaries of their respective disciplines.